\documentclass[sigconf]{acmart}

\AtBeginDocument{%
  \providecommand\BibTeX{{%
    \normalfont B\kern-0.5em{\scshape i\kern-0.25em b}\kern-0.8em\TeX}}}

\acmConference[The Third In2Writing Workshop]{Make sure to enter the correct
  conference title from your rights confirmation emai}{CHI’24}{Honolulu, HI, USA}
\newcommand{\name}[1]{WordDecipher}
\newcommand{\bpstart}[1]{\vspace{1mm} \noindent{\textbf{#1}}.}

\newcommand{\revision}[1]{\textcolor{black}{#1}}
\usepackage{fontawesome5}
\usepackage{subcaption}
\usepackage{stfloats}

\title{WordDecipher: Enhancing Digital Workspace Communication with Explainable AI for Non-native English Speakers}
\author{Yuexi Chen}
\email{ychen151@umd.edu}
\affiliation{%
  \institution{University of Maryland}
  \city{College Park}
  \state{Maryland}
  \country{USA}
}

\author{Zhicheng Liu}
\email{leozcliu@umd.edu}
\affiliation{%
  \institution{University of Maryland}
  \city{College Park}
  \state{Maryland}
  \country{USA}
}

\setcopyright{none}

\renewcommand\footnotetextcopyrightpermission[1]{}

\begin{document}

\begin{abstract}
Non-native English speakers (NNES) face challenges in digital workspace communication (e.g., emails, Slack messages), often inadvertently translating expressions from their native languages, which can lead to awkward or incorrect usage. Current AI-assisted writing tools are equipped with fluency enhancement and rewriting suggestions; however, NNES may struggle to grasp the subtleties among various expressions, making it challenging to choose the one that accurately reflects their intent. Such challenges are exacerbated in high-stake text-based communications, where the absence of non-verbal cues heightens the risk of misinterpretation. By leveraging the latest advancements in large language models (LLM) and word embeddings, we propose \name{}, an explainable AI-assisted writing tool to enhance digital workspace communication for NNES. \name{} not only identifies the perceived social intentions detected in users' writing, but also generates rewriting suggestions aligned with users' intended messages, either numerically or by inferring from users' writing in their native language. Then, \name{} provides an overview of nuances to help NNES make selections. Through a usage scenario, we demonstrate how \name{} can significantly enhance an NNES's ability to communicate her request, showcasing its potential to transform workspace communication for NNES.
\end{abstract}



\begin{teaserfigure}
\includegraphics[width=\textwidth]{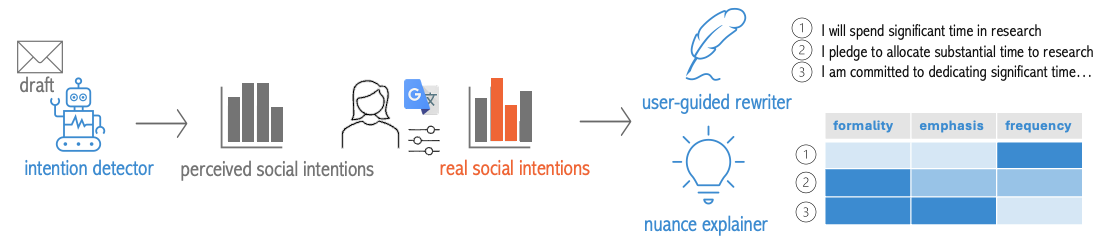}
  \caption{An overview of WordDecipher, an explainable AI writing assistant for non-native English speakers (NNES): upon receiving users' writing, \textit{intention detector} returns perceived social intentions, which could be adjusted by users either numerically or by providing writings in their native language. Then \textit{user-guided rewriter} generates multiple suggestions, and the \textit{nuance explainer} helps NNES make selections.}
  \label{overview}
\end{teaserfigure}

\maketitle

\section{Introduction}
Non-native English speakers (NNES) represent a significant demographic, with more than two million international students enrolled in universities in English-speaking countries~\cite{internationalWiki} and 29.8 million foreign-born workers comprising 18.1\% of the US civilian labor workforce~\cite{Foreignb89:online}. NNES encounter unique challenges in navigating workspace communication, especially text-based communication (e.g., Emails, Slack messages~\cite{slack}), where non-verbal cues are missing. NNES may unintentionally translate phrases from their native language, leading to awkward or incorrect expressions. This increases the risk of miscommunication and decreases the chance of receiving a reply~\cite{lim2022understanding}. 

Some NNES have come up with coping strategies to make the expression sound more natural: they manually verify the existence of phrases in reliable human-authored texts (e.g., New York Times) via tools like Ludwig~\cite{Ludwig, kim2023towards} or general search engines to gauge the prevalence of a phrase based on the volume of returned results~\cite{kim2023towards}. Several AI-assisted writing tools are equipped with rewriting features~\cite{Wordtune31online, quillbot, grammarly, Writefull}, usually providing users with multiple parallel suggestions. Some tools also categorize suggestions by tones~\cite{grammarly, quillbot} and assign fluency scores to each suggestion~\cite{ito2020langsmith, ito2023use}.

However, existing tools are not tailored for NNES. Unlike native English speakers (NES), when facing multiple expressions, NNES may struggle to discern the nuances and select the one that best conveys their intention. Besides, NNES have expressed doubt of the rationales of fluency scores, relying on them for lack of alternatives~\cite{ito2023use}. NNES also expressed the concerns that paraphrasing may alter the original meanings, and the AI-generated suggestions may not fit into their context~\cite{kim2023towards, wegmann2021does, wegmann2022same}.

In response to these challenges, we propose \name{}, an explainable AI-assisted writing tool tailored for non-native English speakers to enhance digital workspace communication. \name{} follows several intent-based design patterns including ``inferring intent'', ``interpretive refraction'', and ``contextualizing choices'' mentioned by Kreminski et al~\cite{kreminski2021reflective}. As shown in Figure~\ref{overview}, \name{} works as follows: \hfill

\bpstart{Intention detector} Given an expression constructed by NNES, \name{} detects several dominant social intentions (e.g., informal-formal, indirect-direct, distant-close~\cite{lim2022understanding}) via large language model (LLM) prompting~\cite{prompt}, then it quantifies the intensity based on pre-trained style embeddings~\cite{patel2023learning, wegmann2021does, wegmann2022same} and displays it to users. 

\bpstart{User-guided rewriter} Should there be a discrepancy between the detected and intended social intention intensities, NNES can either submit a version in their native language to express the intention more accurately or manually adjust the expected social intention intensity score. Upon updating the intensity level, \name{} generates various suggestions by instructing the LLM to revise the text in alignment with the actual intentions.

\bpstart{Nuance explainer} Given parallel suggestions,
\name{} computes their distance in content and style embedding spaces trained on credible human-written resources, and provides an overview of nuances to enable a more informed selection process.

\section{Related work}
\subsection{AI-assisted rewriting tools}
Existing tools like Grammarly~\cite{grammarlyTone}, Quillbot~\cite{quillbot}, WordTune~\cite{Wordtune31online}, and Writefull~\cite{Writefull} offer diverse functionalities in rewriting, supporting various tones, modes, and providing parallel writing suggestions. Boomerang Respondable~\cite{respondable} generates real-time predictive metrics (e.g., word count, reading level, and positivity) for Email response likelihood based on users' writing. LaMPost~\cite{goodman2022lampost} supports rewriting emails based on built-in suggestions (e.g., formal, simpler) and customized instructions. $\mathcal{R}$3~\cite{du2022read} is an iterative text revision system that reduce human efforts by incorporating model-generated revisions and user feedback. WordCraft~\cite{yuan2022wordcraft} supports customized rewrite prompts for creative writing, e.g., ``rewrite this text to be more Dickensian''.
Compared to existing AI-assisted rewriting tools, \name{} is tailored for NNES, incorporating explainable AI features to facilitate comparing and selecting parallel suggestions.

\subsection{Non-native English speakers (NNES) research}
Studies have explored NNES interactions with AI rewriting tools, highlighting challenges in accurate message delivery and paraphrase evaluation. For instance, Ito et al.~\cite{ito2023use} found that NNES struggled with evaluating AI-generated output, often relying on machine translations. However, machine-translated emails from NNES tend to convey social intentions less accurately compared to those written directly in English by NNES~\cite{lim2022understanding}. NNES are also found to accept more Email writing suggestions than NES, potentially due to difficulties in identifying issues~\cite{buschek2021impact}. Comuniqa~\cite{mhasakar2024comuniqa} aims to enhance NNES' speaking skills through LLM-based evaluations across multiple dimensions. Kim et al.~\cite{kim2023towards} examined NNES' coping strategies, including referring to human-authored texts from credible sources, seeking textual explanations and statistical evidence via search engines. \name{} draws inspiration from previous studies with NNES, and possesses more explainable features and tailored support for NNES.

\subsection{Pre-trained text representations}
Word embeddings transform words into vector representations that reflect their context from large datasets~\cite{devlin2018bert, liu2019roberta}, enabling the derivation of sentence or document embeddings through aggregation~\cite{reimers2019sentence, opitz2022sbert}. Such embeddings are instrumental for downstream tasks like computing text similarity. However,
the same content can be conveyed in different styles, e.g., ``r u a fan of them or something?'' vs. ``Are you one of their fans?''~\cite{wegmann2021does}. Addressing this, Wegmann et al. ~\cite{wegmann2022same} trained style-specific embeddings by controlling the content and fine-tuning RoBERTa models~\cite{liu2019roberta}. Recently, Patel et al.~\cite{patel2023learning} harnessed large language models (LLM) to annotate styles and synthesize a large training dataset, leading to the creation of LISA embeddings, 768-dimensional interpretable style vectors designed for interpretability, enabling style analysis within a multidimensional framework. \name{} is partly empowered by pre-trained embeddings to conduct quantitative comparisons in content and styles.

\section{Usage scenario}
This section illustrates \name{}'s application through Yinuo's experience. Yinuo, a sophomore chemistry major from China attending a public university in the U.S., has lived in the country for one and a half years. Last semester, Yinuo took a class taught by Prof. Miller and was captivated by the latest advancements in nanotechnology, so Yinuo decided to ask whether it's possible to join his lab as an undergrad research assistants in the summer. 

\begin{figure*}
    \centering
    \includegraphics[width=0.8\textwidth]{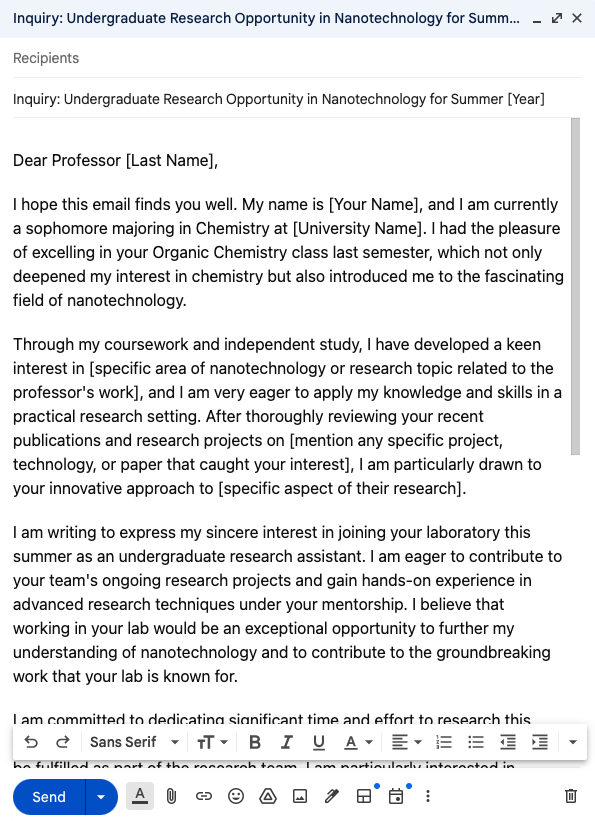}
    \caption{ChatGPT-generated email for Yinuo's inquiry on research opportunities in Prof. Miller's lab}
    \label{fig:yinuo}
\end{figure*}

First, Yinuo begins by prompting ChatGPT~\cite{chatGPT}: ``I'm a sophomore chemistry major, last semester I took an organic chemistry class and got an A. I'm really interested in nanotechnology; help me write an email to tell him that I want to join his lab in the summer.'' Upon receiving ChatGPT's reply, Yinuo scans the content and believes it is a good starting point, so she pastes it into her school's Gmail interface, as shown in Figure~\ref{fig:yinuo}. However, she notes the email's length and formal tone, which do not align with Prof. Miller's casual demeanor, yet she wishes to maintain a respectful tone appropriate to her cultural background. She makes revisions manually and asks follow-up questions to ChatGPT, but still does not feel confident enough to send out the email. Then, she asks whether her friends have written similar emails, and one friend recommends \name{} to her. 

\name{} takes a draft as input (Figure~\ref{overview}), so Yinuo pastes her email draft there. For Yinuo's email, \name{} detects several social intentions: respectful-disrespectful, formal-informal, distant-close, shy-boldness, and assigns a score on a 7-point scale to each dimension. Yinuo notices the discrepancy between the perceived and her real intentions. She thinks Prof. Miller should remember her as she sometimes asked him questions after class, so she adjusts the score towards more ``informal'' and ``close'', but keeps ``respectful'' the same. After updating the scores, \name{} generates several rewriting suggestions on the paragraph level (also supporting sentence or word-level granularity). \revision{\name{} also explains nuances of parallel suggestions to help Yinuo quickly grasp the subtleties, which makes Yinuo feel much more confident to send out the email.}

Next time, Yinuo plans to recommend \name{} to her cousin Muchen in China, \revision{who recently landed a job in a multinational company. Muchen is less proficient in English than Yinuo, so for critical work-related communication, Muchen initially drafts in Chinese before translating into English. This process may distort the intensities of social intentions~\cite{lim2022understanding}, and WordDecipher can assist Muchen to accurately convey the intended messages.}

\bibliographystyle{ACM-Reference-Format}
\bibliography{main}

\end{document}